# Fast neutron spectroscopy from 1 MeV up to 15 MeV with Mimac-FastN, a mobile and directional fast neutron spectrometer

N. Sauzet [*], D. Santos, O. Guillaudin, G. Bosson, J. Bouvier, T. Descombes, M. Marton, JF. Muraz

*Laboratoire de Physique Subatomique et de Cosmologie (LPSC) – Université Grenoble Alpes - CNRS/IN2P3 – 53, avenue des Martyrs, 38026 GRENOBLE cedex – France*

**Abstract**

In the frame of direct dark matter search, the fast neutrons producing elastic collisions are the ultimate background. The MIMAC (MIcro-tpc MAtrix Chambers) project has developed a directional detector providing the directional signature to discriminate them based on 3D nuclear tracks reconstruction. The MIMAC team of the LPSC has adapted one MIMAC chamber as a portable fast neutron spectrometer, the Mimac-FastN detector, having a very large neutron energy range (10 keV – 600 MeV) with different gas mixtures and pressures. The present paper shows its main features and functionality and demonstrates its potential in the energy range from 1 MeV to 15 MeV at the GENESIS neutron source facility of LPSC.

## 1   Introduction

In the frame of direct dark matter detection, the fast neutrons producing elastic collisions on nuclei in the active volume of detection generate the same signals that an eventual WIMP (Weakly Interacting Massive Particle). These events, as well as those produced by neutrinos, are the ultimate background for dark matter detection. In the frame of the MIMAC project (MIcro-tpc MAtrix Chambers), a directional detector giving the directional signature to discriminate them has been developed [ref 1]. The LPSC MIMAC team has adapted one MIMAC chamber as a fast neutron spectrometer. Fast neutron spectroscopy is a detection challenge and required in many different domains, such as neutron dosimetry, identification of special nuclear material and nuclear physics. Some applications require measurements above 10 MeV and on a large energy range. Among these applications, we can mention the secondary neutrons in radiotherapy, and the characterization of atmospheric neutrons produced by cosmic particles, going up to 100 MeV.

In general, neutron spectroscopy at high energies (above 1 MeV) is challenging for the present available detector technologies. While, technologies exist for neutron spectroscopy, the more widely used technologies have some limitations. For example, neutron energies can be accurately measured by time of flight, but such measurements require time difference measurements between the neutron emission from a source and its detection. If the source is unknown, the kinetic energy cannot be obtained. Detection using neutron capture after thermalization is another approach, but it often leads to poor energy resolution and requires a prior, selecting the expected neutron energies. The detection in solids through elastic collisions is limited due to the absorption of recoils in the converter, whereas detection in liquid scintillators results in a limited measuring range. Developments of directional fast neutron detectors using Time Projection Chamber (TPC) filled with $H_2$ are focusing on the detection of special nuclear material (SNM) and the localization of fission neutron source, which results in a limited neutron energy range [ref.2 and ref.3]. These detectors are currently non-mobile detectors.

In the present paper, we describe a fast neutron spectrometer called Mimac-FastN, that tackles these issues for high neutron energies, based on the 3D detection of nuclear recoils with a very fast sampled, self-triggered and low noise electronics developed at the LPSC [ref.4 and 5]. The main goal of the paper is to describe the instrument features, to give an overview of the performance, and to demonstrate the promising perspectives opened, through the presentation of measurements done in mono-energetic neutron fields. In particular, we present an instrument set-up to cover a large neutron energy range, from 200 keV to more than 15 MeV, with a design strategy focused on a mobile device without flammable gas or regulated matters, suitable for industrial areas. A non-mobile version of Mimac-FastN, the µTPC LNE-IRSN-MIMAC prototype, has been the subject of two PhD-thesis [ref.6 and 7], in the frame of a neutron metrology collaboration.

The first part of this work is dedicated to the detection principle description followed by the facility presentation where the mono-energetic measurements have been performed. The gas mixture specificity that gives access to high-energy neutrons measurements is presented. In the following section, a focus is made on the data analysis strategy to reconstruct the neutron energy. The future additional work to achieve final neutron spectroscopy quality measurements with this instrument is discussed in the end of the paper.

## 2    Detection principle

Mimac-FastN is a micro-TPC (Time Projection Chamber) based on a micro-pattern detector coupled to a fast self-triggered electronics [ref.1]. The chamber is filled with a gas mixture being itself the converter of fast neutrons into charged particles. In the present paper, we describe the operation of Mimac-FastN with 2 liters of a gas mixture of 95 % of $^4$He and 5 % of $CO_2$ as a quencher, at 700 mbar, to measure neutron energies between 1 MeV up to 15 MeV. The gas mixture and the pressure can be adapted to the envisaged application energy range [ref.8] and [ref.9]. Fast neutron detection is performed through the 3D tracking of the nuclear recoils resulting from elastic scattering between incident fast neutrons and the gas nuclei. The nuclear recoils stopping in the active volume lose only part of their kinetic energy by ionization in the gas volume. The primary electrons resulting from this ionization process are collected by an electrical field of 160 V/cm through a 25 cm long drift chamber, up to the micro-pattern detector (a square bulk Micromegas [ref.10] with a 512 µm gap, and sides of 10.8 cm). A high electrical field of 10.5 kV/cm between the grid and the anode of the Micromegas produces avalanches, giving the signal amplification. Resulting secondary electrons are collected on the pixelated anode and the positive ions drifting towards the grid.

The Mimac-FastN electronic board [ref.4] manages two synchronized types of data. The first one is the energy released in ionization by the nuclear recoil, read through a charge preamplifier connected to the mesh of the micro-pattern detector. This preamplifier, developed by the LPSC, has a gain of about 100 mV/pC, adjustable depending on the energy range required and a time constant of 2 ms. In such a way the rise time of the signal is small compared to the electronic decay time. The second type of data is the fired strips of pixels on the anode of the micro-pattern detector (512 strips, 256 on X and 256 on Y), which gives access to the 2D position projection of the primary electron cloud.

The data on the grid and on the pixelated anode are read out at a sampling frequency of 40 or 50 MHz, depending on the length of tracks to be produced, and managed by the electronic board. In this way, each nuclear recoil track is sliced in samples. For each sample, the barycenter of the X and Y coordinates is calculated from the fired pixels on the anode. The neutron energy measurement relies on the definition of the nuclear recoil track direction. Two parameters play a major role on this definition: the time sampling and the electronic noise. So a fast time sampling gives access to a better resolution on this determination of the track step position, and finally on the track direction. In a similar way, the lower the electronic noise on the strips signal is, the higher the spatial resolution is on the barycenter of fired pixels.

In the gas mixture $^4$He/$CO_2$ (5%) at 700 mbar and at 40 MHz sampling, each sample has a perpendicular component to the anode of 241 µm (referring to a Magboltz simulation that gives a drift velocity of 9.65 µm/ns in this gas mixture, which leads to a length of 9.65 µm/ns x 25 ns). So the 3D nuclear recoil track is reconstructed by the composition of the 2D picture on the pixelated anode, and the perpendicular component (delta Z) inferred from the electronic sampling. The absolute Z coordinate is not measured, and the assumption is made that the elastic scattering interaction has occurred in the middle of the drift cage. This assumption introduces an error on the mean kinetic energy of the neutron, estimated by Geant4 simulations at 0.4 % for neutrons of 3 MeV, and at 4% at 15 MeV.

The electronic board is coupled to the micro-pattern detector through an interface board that ensures the chamber tightness. This very low noise electronic board manages itself the triggering of each event acquisition through a FPGA. The acquisition triggering is done from the signal on the grid requiring an ionization energy threshold. Once triggered, the acquisition window remains open 25 µs at maximum.

The synchronization of the readout on the grid (ionization energy) and the readout on the pixelated anode (track data) is managed by the FPGA. The sampling time is the same for the pixelated anode reading, and for the charge profile on the grid. The two different types of information being the track coordinates and the deposited charges can then be synchronized for each time-slice.

The pixelated anode readout is performed by the 8 MIMAC ASICs, specifically developed by the MIMAC team of the LPSC [ref. 5].

A dedicated software controls the FPGA through a USB connection, and stores the data event by event in a text file or in a PostgreSQL database.

Analysing the event-by-event sampled data from the grid and the pixelated anode, the kinetic energy of the incident neutron can be measured.

The neutron kinetic energy is deduced from the kinetic energy of the nuclear recoil by the kinetic energy conservation in elastic scattering giving the following equation :

$$E_n = \frac{(1+m_R)^2}{4m_R} \times \frac{E_R}{cos^2(\theta_{RN})}$$

being $E_n$ the incident neutron energy, $E_R$ the kinetic energy of the nuclear recoil, $\theta_{RN}$ the angle between the nuclear recoil track and the incident neutron direction, and $m_R$ the nuclear recoil mass (in GeV/c²). The neutron mass is approximated to 1 GeV/c².

The kinetic energy of the nuclear recoil is determined from the measured ionization energy, affected by the ionization quenching factor ([ref.11] and [ref.12]) in the considered gas mixture.

The angle $\theta_{RN}$ is estimated from the 3D recoil track reconstruction, taking into account the neutron emitter position giving the incident neutron direction (see Figure 1).

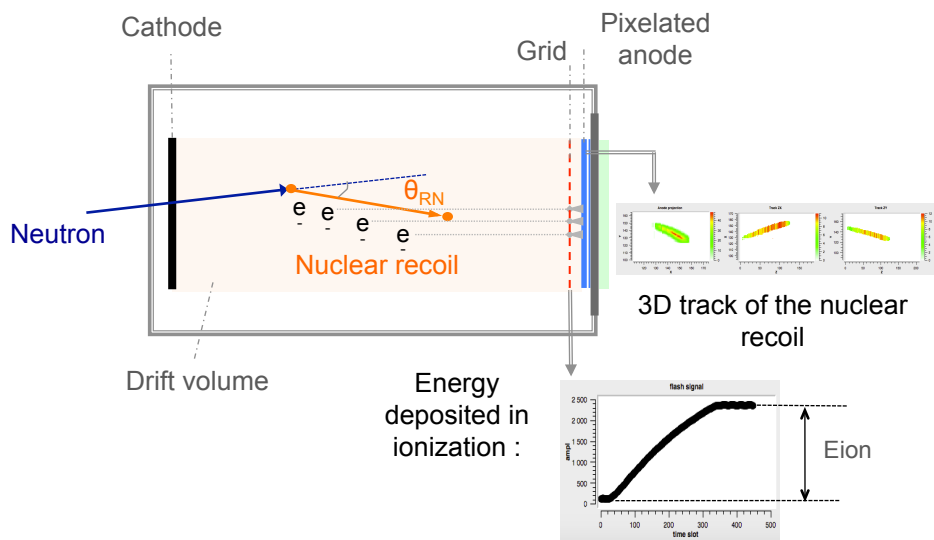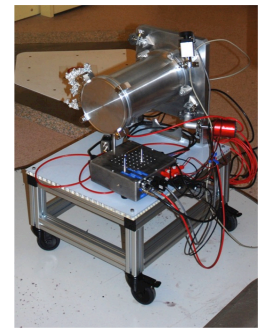

Figure 1 : A schematic drawing of the detector structure at the left, and picture of Mimac-FastN at the right.

*See online version for colors*

## 3   GENESIS facility

Spectrometry of the neutrons produced by the reactions D(d(220 keV,n) and T(d(220 keV,n) has been explored at the GENESIS facility [ref.13], with Mimac-FastN, for neutrons mono-energetic measurements at 3 MeV and 15 MeV. A picture of the facility with the experimental set-up can be seen in Figure 2, enclosed by its concrete bunker.

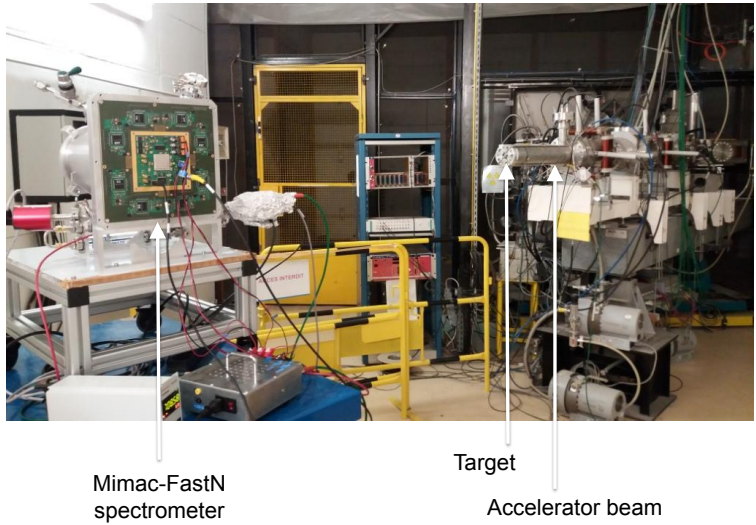

Mimac-FastN spectrometer

Target

Accelerator beam

Figure 2 : Experimental set-up at GENESIS, for the measurement in the perpendicular configuration of Mimac-FastN, with the reaction T(d(220 keV),n).

*See online version for colors*

## 4  Special features of nuclear elastic collisions with neutrons above 3 MeV

The angular distribution of the $^4$He recoils, resulting from elastic scattering with fast neutrons, is a function of the neutron energy. Angular distributions in the laboratory frame have been calculated with the Monte Carlo code Geant4 [ref. 14], version 10.1.2, with the physics list QGSP_BERT_HP_LIV, considering a chamber filled with a gas mixture of $^4$He/$CO_2$ (5%) at 700 mbar, see the results for three different neutron energies in Figure 3.

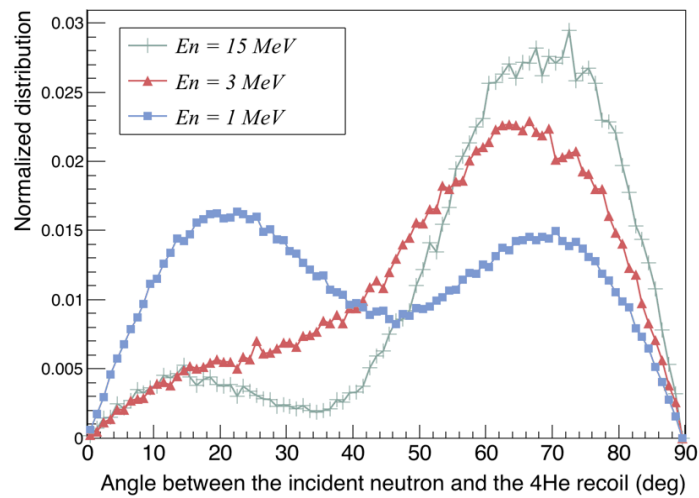

Figure 3 : Angular distributions of $^4$He recoils, resulting from elastic scattering with neutrons of 1 MeV, 3 MeV and 15 MeV.

*See online version for colors*

These angular distributions show that for neutron energies above 3 MeV, angles between 55° and 85° are more likely. This has two implications. The first one is that the most probable kinetic energy of the $^4$He recoil is 1.1 MeV for a neutron of 15 MeV. In a mixture of $^4$He/$CO_2$ (5%) at 700 mbar, a recoil track with this kinetic energy is 3.3 cm long according to SRIM [ref. 15]. This track length is far shorter than the drift chamber length described previously, and so this track has a high probability to remain contained in the volume. From these angular distribution simulations, we deduce that above 3 MeV, the higher the neutron energy is, the higher the elastic scattering angular is, then the lower the energy carried away by the $^4$He recoil is, and then the smaller the recoil track mean length will be.

The second implication is that a 3D geometry is needed to get kinetic energies of the neutrons from detected nuclear recoils, for neutron energies above 3 MeV. The advantage of a gaseous detector like Mimac-FastN with a cylindric or cubic symmetry geometry is that nuclear recoils can be detected whatever their direction is. If the nuclear recoil tracks are parallel to the pixelated anode, the 3$^{rd}$ dimension calculated by the electronic sampling will be limited to a few samples. However, knowing the position of the neutron emitter, the chamber can be orientated perpendicularly to the emission direction. In such a configuration, if the $\theta_{RN}$ angles are above 60°, the tracks' orientations are optimized compared with the anode plane (see Figure 4).

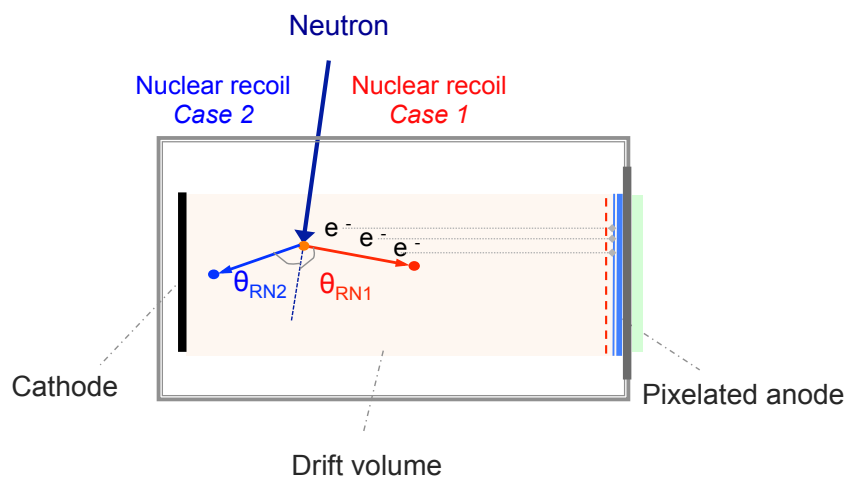

Figure 4 : Measure configuration for neutron energies above 3 MeV.
*See online version for colors*

In this perpendicular configuration, two groups of nuclear recoils can be differentiated: recoils directed towards the cathode, and recoils directed towards the anode. Having the head-tail signature of the nuclear recoil tracks is a prerequisite to the neutron energy calculation and diffused neutron discrimination. In order to determine the direction of a nuclear recoil track by means of the acquired data, a new observable needs to be defined, see section 5.6.

## 5 Data analysis strategy

### 5.1 Principle of data analysis

In the following subsections, we describe the different steps to reconstruct the neutron energies from the detected nuclear recoils, event by event. The first prerequisite is the energy calibration of the Flash-ADC. Then, we define some observables to identify $^4$He recoils from other nuclear recoil species and gamma rays. The selection of the nuclear recoils related to the neutron energy reconstruction, starts discriminating all the tracks that are not completely contained within the drift chamber, then with the selection of those $^4$He recoils along with the attribution of their 3D direction, and finally with the discrimination of the recoils resulting from the interaction of scattered neutrons not coming directly from the neutron emitter.

### 5.2 Energy calibration

The charge profile is measured through a charge preamplifier connected to the mesh and sent to a Flash-ADC (on the electronic board) that digitizes the signal on 4096 channels. The Flash-ADC energy calibration is done through a natural boron coating fixed on the cathode and by means of the following capture reaction of thermal neutrons by the $^{10}$B isotope:

$$n + {}^{10}_{5}B \rightarrow {}^{4}_{2}He + {}^{7}_{3}Li + \gamma \quad (94\ \%)$$

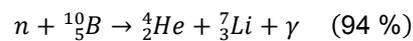

$$n + {}^{10}_{5}B \rightarrow {}^{4}_{2}He + {}^{7}_{3}Li \quad (6\ \%)$$

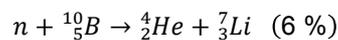

The boron coating consists in an IBS (Ion Beam Sputtering) deposit of 500 nm of $^{nat}B_4C$ on an aluminum sheet (see picture in Figure 5). The $^{10}$B isotope represents 20% of the natural boron. The coating has a specific shape in order to check the spatial resolution of the boron projection picture on the anode.

The energies deposited in ionization by the $^4$He and $^7$Li particles are measured on the Flash-ADC, and their tracks are imaged on the pixelated anode.

Figure 5 shows the anode projection of the detector exposed to a 3 MeV neutron field crossing 5 cm of high density polyethylene. The projection on the anode of the first point of all the tracks directed towards the anode, highlights the boron coating, due to neutron capture, that become predominant with respect to the total elastic scattering on the gas nuclei, at low energies. The presence on this picture of the specific shape of the boron corner proves the uniformity of the electrical field lines in the field cage and gives an estimation of the spatial resolution.

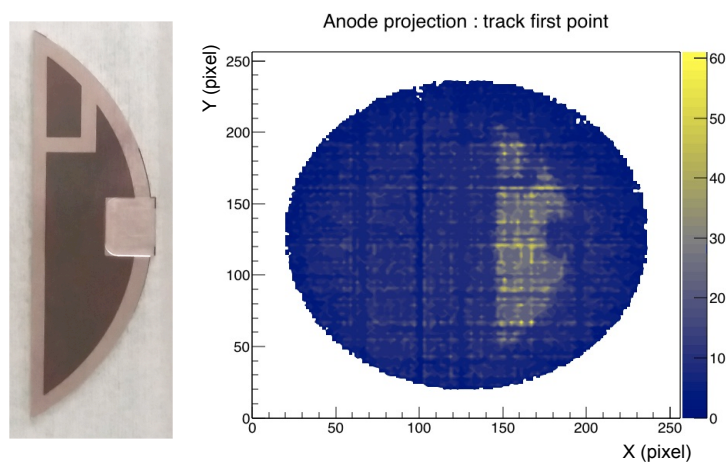

Figure 5 : on the left, picture of the B4C coating used for energy calibration and spatial resolution measurements ; on the right, projection on the pixelated anode of the first point of all the tracks, that highlights the boron shaped coating, in a neutron beam of 3 MeV moderated through 5 cm of HDPE.

*See online version for colors*

A selection of all the tracks whose interaction points are located on the boron coating projection, leads to the Flash-ADC spectrum of the particles issued from neutron capture on $^{10}$B presented in Figure 6. In order to convert the ADC channels in energy units (keV), this measured spectrum is compared to a first order polynomial calibration equation minimizing the difference between the resulting measured spectrum and the ionization energy spectrum calculated with Geant4, shown in Figure 7. This minimizing process gives the two linear calibration parameters.

The peak issued from the alpha particle emitted in the case of a branching ratio of 6 % constitutes a checkpoint of the calibration equation.

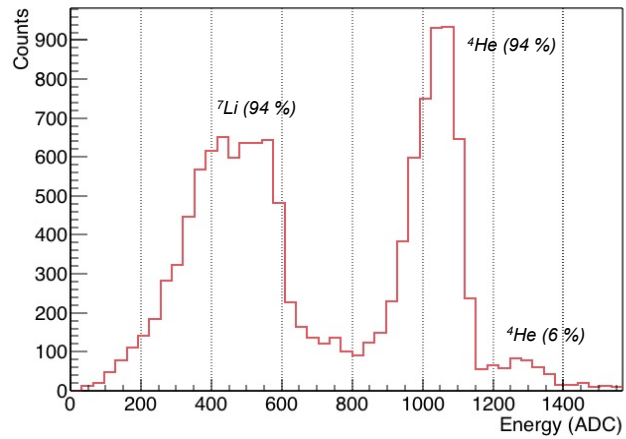

Figure 6 : Measured ionization energy spectrum of the $^4$He and $^7$Li particles resulting from neutron capture on the boron coating, in a neutron beam of 3 MeV moderated through 5 cm of HDPE. This spectrum is the raw data from the Flash-ADC.

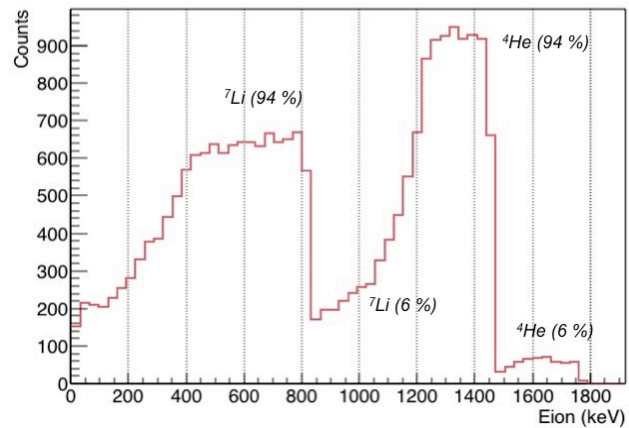

Figure 7 : Simulated ionization energy spectrum of the $^4$He and $^7$Li particles resulting from neutron capture on the boron coating, taking into account the coating thickness, and the Geant4 ionization quenching factor.

The Geant4 simulation embeds an ionization quenching factor, which is defined as the amount of ionization energy deposited by the particle, compared to its kinetic energy ([ref. 11], and [ref. 12]). In the energy ranges considered above, the mean ionization quenching factor considered by the Geant4 simulation is 97,5 % for $^7$Li, and 98,8 % for $^4$He. We have recently conducted measurements of the ionization quenching factors of $^4$He and protons in the gas mixture proposed. More details on these measurements and on the energy calibration method will be given in a future paper dedicated to uncertainties analysis.

### 5.3 Discrimination of species and physical process

Elastic scattering can occur on all the gas nuclei constituting the gas chamber, so recoils of $^4$He, $^{12}$C and $^{16}$O are detected. Besides, fast neutrons can interact with all the structures being part of the detector, mostly composed of aluminum, Kapton, PMMA and copper. The first step will be to list all these possible interactions with the detector structure. The second step will be to describe a way to discriminate the $^4$He recoils from all the other species.

#### 5.3.1 Interactions on the detector structure

##### 5.3.1.1 *Fast neutrons on aluminum*

Aluminum is the material of the ionization chamber. On $^{27}$Al, fast neutrons can produce (n,p) and (n,α) reactions, respectively above 1.8 MeV and 3.1 MeV. For neutrons of 15 MeV, the cross-sections of these reactions are roughly 0.06 barn and 0.1 barn respectively, and protons are emitted with an energy up to 12.7 MeV, while alphas with an energy up to 10.2 MeV can be produced. These alpha particles are all absorbed by the materials crossed before reaching the gaseous active volume (cathode and field cage), whereas protons passing through releasing a residual energy can be detected in the active volume.

##### 5.3.1.2 *Fast neutrons on Kapton and PMMA*

Kapton and PMMA are components of the field cage that define the drift field and the active gaseous volume.

On Kapton and PMMA, fast neutrons can interact through elastic scattering with hydrogen, or through inelastic (n,p) and (n,α) reactions.

Elastic scattering on hydrogen has a cross section of 0.6 barn for neutrons of 15 MeV.

(n,p) reactions mainly occur on $^{16}$O and $^{14}$N for neutron energies above 9.6 MeV and 0.6 MeV respectively. For neutrons of 15 MeV, cross sections of such reactions are roughly of 0.04 barn, and protons are emitted with energies up to 5.08 MeV and 13.4 MeV. Then, proton recoils or protons issued from nuclear reactions can be detected in the gaseous active volume.

(n, α) reactions mainly occur on $^{12}$C, $^{16}$O and $^{14}$N, for neutron energies above 5.7 MeV, 2.2 MeV and 0.2 MeV respectively. For neutrons of 15 MeV, cross sections of such reactions are respectively of 0.05 barn, 0.14 barn and 0.08 barn, and alpha particles are released with energies up to 6.4 MeV, 9.8 MeV and 10.9 MeV. These alpha particles issued from nuclear reactions can be detected in the gaseous active volume.

All these particles leave some energy in the Kapton or the PMMA itself, before being released in the gaseous active volume. The specificity of all these particles is that they are emitted on the edges of the active volume, and as such, they can be discriminated by analysis of their projected track on the pixelated anode.

### 5.3.1.3 *Fast neutrons on copper*

Copper is the cathode material as well as a component of the field cage.

On $^{63}$Cu (69 % of natural copper), fast neutrons can produce inelastic (n,p) and (n,α) reactions, above 0.7 MeV and 1.7 MeV respectively. For neutrons of 15 MeV, the cross-sections of these reactions are 0.06 barn and 0.04 barn respectively, and protons are emitted with an energy up to 14 MeV, while alphas with an energy up to 12.5 MeV. All these particles can release their energy partially or totally in the active volume.

*5.3.1.4  Summary of the main interactions with the detector structures, for neutrons of 15 MeV*

The table hereafter is a summary of the main interactions when the chamber is placed in a neutron field of 15 MeV.

| Interaction process | Material | Cross-section | Max energy of the product |
|---|---|---|---|
| (n,p) | Aluminium | 0.06 barn | 12.7 MeV |
| $^1$H(n,el) | Kapton, PMMA | 0.6 barns | 15 MeV |
| (n,p) | Kapton, PMMA | 0.04 barn | 5.08 MeV / 13.4 MeV |
| (n,α) | Kapton, PMMA | 0.05 barn | 6.4 MeV |
|  |  | 0.14 barn | 9.8 MeV |
|  |  | 0.08 barn | 10.9 MeV |
| (n,p) | Cuivre | 0.06 barn | 14 MeV |
| (n,α) | Cuivre | 0.04 barn | 12.5 MeV |

The major contributors come from the reaction $^1$H(n,el) on Kapton and PMMA.

### 5.3.2  Track length as a function of ionization energy

In our approach to calculate the neutron energies, we consider the $^4$He recoils only. So all the other recoil species and the contributions of nuclear reactions have to be discriminated. In order to explore the discrimination of the different species, giving the electronic synchronization of the pixelated anode with the grid, we can plot each event track length as a function of its energy released in ionization in the gaseous active volume. We performed both simulations and measurements of these distributions, as shown in the next paragraphs.

*5.3.2.1 Geant4 Simulations*

The two-dimensional graph (track length as a function of ionization energy) gives the possibility to discriminate the different physical process, as can be assessed by the result of Geant4 simulations of Mimac-FastN in a mono-energetic neutron field of 15 MeV, as shown in Figure 8 and Figure 10.

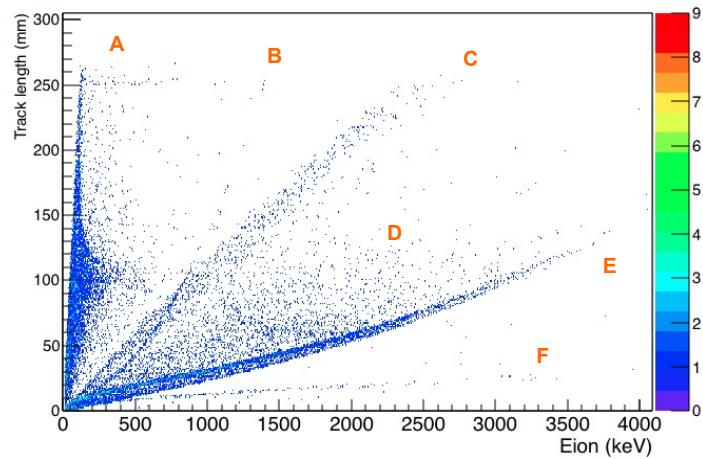

Figure 8 : Simulation of the track lengths as a function of ionization energy, with Mimac-FastN modeled with Geant4, in a configuration parallel to the mean neutron emission, with neutrons of 15 MeV.

*See online version for colors*

On this two-dimensional plot resulting from simulation, different structures emerge that are assigned as follows:

**Branch A** : protons issued from Kapton, PMMA, aluminum and copper, that do not release all their energy in the active volume, because their track lengths are much longer than the distance between the interaction point and the field cage. This branch is broad because the length related to the energy released depends on the initial energy of the proton and the Bragg peak position.

**Branch B** : protons issued from Kapton and PMMA, releasing all their energy inside the active volume.

**Branch C :** $^4$He recoils with a kinetic energy of more than 7 MeV that are scattered in a head-on collision along the longitudinal chamber axis (perpendicular to the pixelated anode), that release little energy in the gaseous volume with long path length, since the stopping power is small at the beginning of the recoil travel inside the gas (the Bragg peak is at the end of the track). See Figure 9 for a focus on these $^4$He recoils.

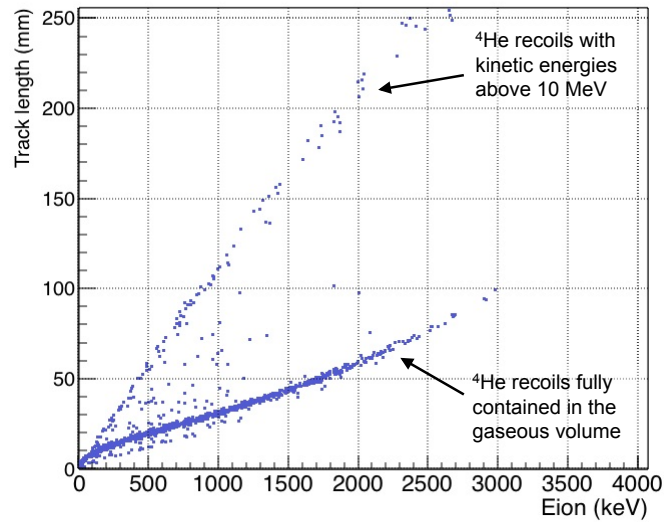

Figure 9 : Focus on the simulation of the track lengths of $^4$He recoils as a function of ionization energy, in a configuration parallel to the mean beam emission, with neutrons of 15 MeV.

**Structure D** : $^4$He recoils of low energy that release part of their energy in the active volume before going outside.

**Branch E** : $^4$He recoils that release all their energy in the active volume. The dispersion on this branch thickness depends on the proportion of $^4$He recoils releasing little amounts of energy outside of the active volume.

**Branch F** : $^{12}$C and $^{16}$O recoils that release all their energy in the active volume.

Given the good separation of all these branches characterizing different physical process and nuclear recoil masses, we can easily select the branch E ($^4$He recoils) that is the main branch of interest for neutron energy measurements discriminating all the other ones. Only the 3D track reconstruction associated with the measurement of its energy released in ionization gives the possibility to discriminate them.

The two-dimensional plot of Figure 10 in the perpendicular configuration shows that the structure with multiple branches associated with different physical process is similar to the one obtained in the parallel configuration. The main difference concerns the branch **C** that is not observed in the perpendicular configuration: the $^4$He recoils scattered with a small angle go outside of the active volume in parallel to the anode, and are then discriminated. In the parallel configuration, the track lengths are longer, which is consistent with the size of the field cage (10.8 x 10.8 x 25 cm).

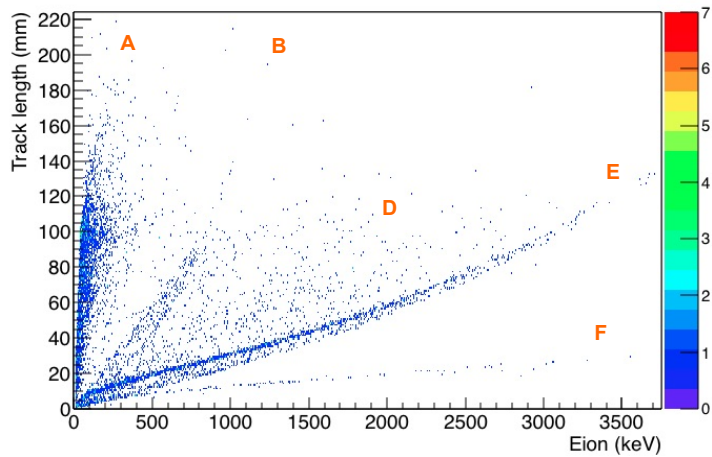

Figure 10 : Simulation of the track lengths as a function of ionization energy, with Mimac-FastN modeled with Geant4, in a configuration perpendicular to the mean beam emission, with neutrons of 15 MeV.

*See online version for colors*

Simulations show that by means of 3D tracks synchronized with the ionization energy measurement, the neutron spectrometer Mimac-FastN presents an excellent ability to identify and discriminate different nuclear recoil species even those that do not remain fully contained inside the ionization active volume.

*5.3.2.2 Measurements with 15 MeV neutrons*

Measurements have been performed with Mimac-FastN at the GENESIS facility [ref.13] with the reaction T(d(220 keV), n), that results in the production of neutrons of 15 MeV at 0°. The set-up of the experiment is described in § 6 and 7.

Figure 11 shows the measured track lengths as a function of ionization energy for all the events detected. We observe on this plot the same structure as the one obtained by simulation, which allows the attribution of the measured branches to the physical process and species. The main $^4$He recoil branch is easily differentiable (branch E).

We have performed a measurement in the same conditions, with the chamber rotated of 90°, and so placed perpendicularly to the mean beam direction. Figure 12 shows the measured track lengths as a function of ionization energy in this second configuration.

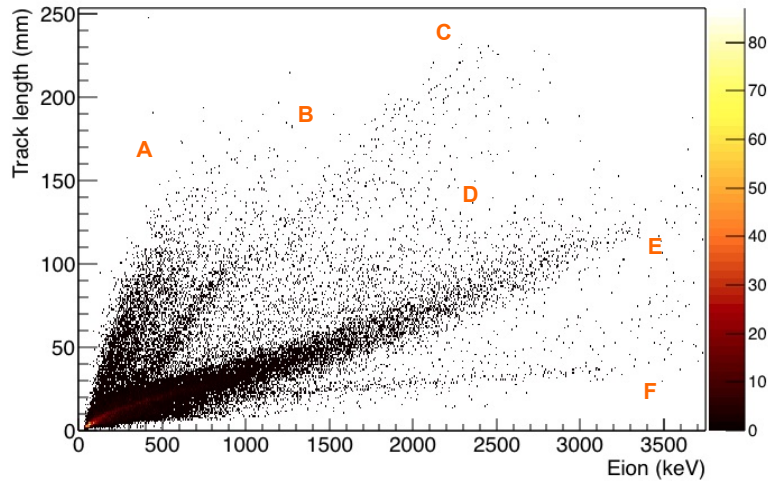

Figure 11 : Measured track lengths as a function of energy, in a configuration parallel to a neutron beam of 15 MeV produced by the reaction T(d(220 keV),n) at GENESIS facility.

*See online version for colors*

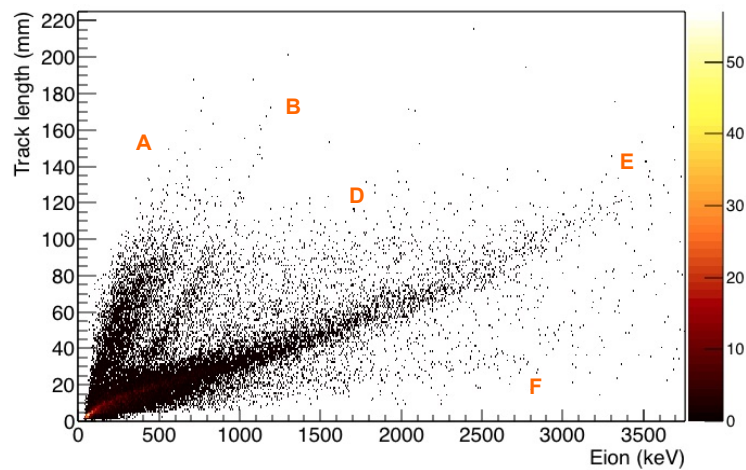

Figure 12 : Measured track lengths as a function of energy, in a configuration perpendicular to a neutron beam of 15 MeV produced by the reaction T(d(220 keV),n) on GENESIS, for tracks directed toward the anode or the cathode.

*See online version for colors*

These two plots corresponding to parallel and perpendicular configurations respectively to the neutron beam, show the same branch structure. Hence, all the structures highlighted by the simulated plots are confirmed by measurements, which prove a good track reconstruction whatever the orientation of the detector with respect to the mean beam axis.

The measured events showed on branch A attributed as the energetic protons going outside the active volume have a particularity compared to $^4$He recoils tracks on branch E: the tracks on the pixelated anode present many holes, whereas $^4$He tracks do not show this type of feature, as seen in Figure 13 for protons and Figure 14 for $^4$He, showing dense and clear tracks, meaning tracks with a low number of holes by time-slice. These holes are due to the low ionization energy deposited per time-slice (on average, 1.5 keV/25 ns (time-slice) for a proton of 400 keV) that is not enough to trigger all the strips of pixels that have a threshold defined individually as a function of their intrinsic noise. For comparison, a $^4$He of 400 keV deposits 8.5 keV/25 ns (time-slice) on average.

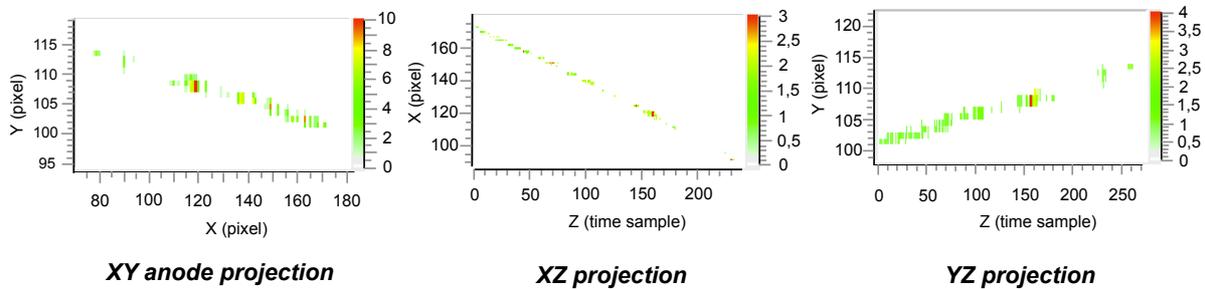

**XY anode projection**  **XZ projection**  **YZ projection**

Figure 13 : A 400 keV proton recoil track issued from an elastic scattering with a neutron of 15 MeV produced by the reaction T(d(220 keV),n) on GENESIS.

*See online version for colors*

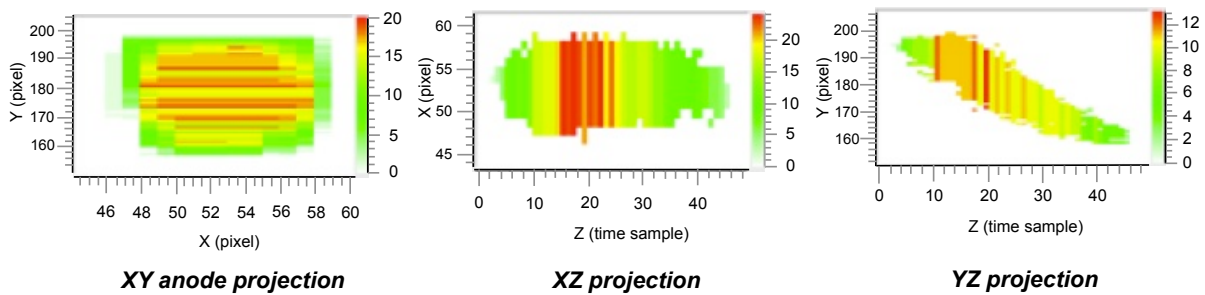

**XY anode projection**  **XZ projection**  **YZ projection**

Figure 14 : A 400 keV $^4$He recoil track issued from an elastic scattering with a neutron of 15 MeV produced by the reaction T(d(220 keV),n) at the GENESIS facility.

*See online version for colors*

### 5.3.3 Gamma ray – neutron discrimination

Compton electrons resulting from the interaction of few MeV gamma rays with the detector structures lose 70 keV at most in the Mimac-FastN gaseous active volume, as attested by a simulation with Geant4 (see Figure 15).

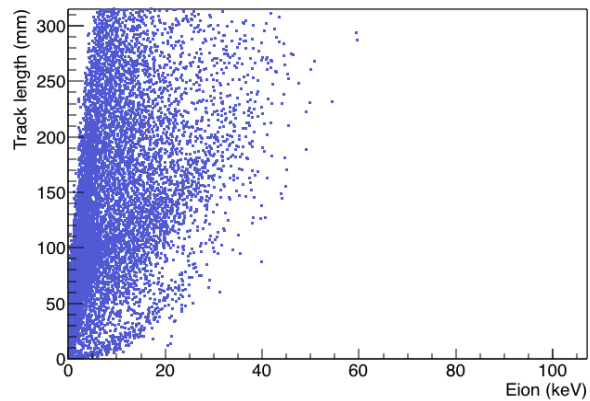

Figure 15 : Simulation with Geant4 of the track lengths of electrons as a function of their ionization energy deposited in the active volume, resulting from the interaction of 100 keV to 15 MeV gamma rays with Mimac-FastN structures.

The minimum ionization energy that can be detected is determined by the noise level on the grid of the micro-pattern detector. This lowest ionization energy threshold is 25 keV with the usual Mimac-FastN set-up. So referring to simulation, above 25 keV, we only have a residual of 0.3 % of electrons. The energy loss per time-slice of these electrons is so low (0.06 keV/time-slice on average) that measurements lead to a similar phenomenon as the one observed for protons : the energy released per unit length by these electrons is not enough to trigger the strips of pixels, and so these events do not leave dense and clear tracks in the active volume, showing only sparse clusters of fired pixels, being easily discriminated. However, the energy of all these events is measured on the grid of the micro-pattern detector, as it can be see in Figure 16 showing a Flash-ADC profile, which has a typical structure of charge clusters, compared to the energy profile of a nuclear recoil showing only one well defined cluster (Figure 17).

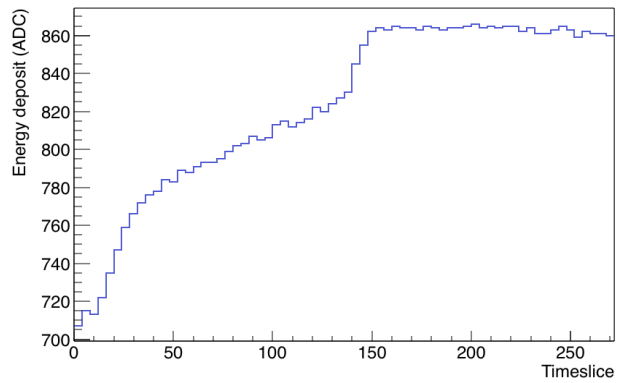

Figure 16 : A 53 keV electron ionization energy release profile measure, resulting from the interaction of a secondary gamma-ray, produced with Mimac-FastN in a field with the reaction T(d(220 keV),n) on GENESIS.

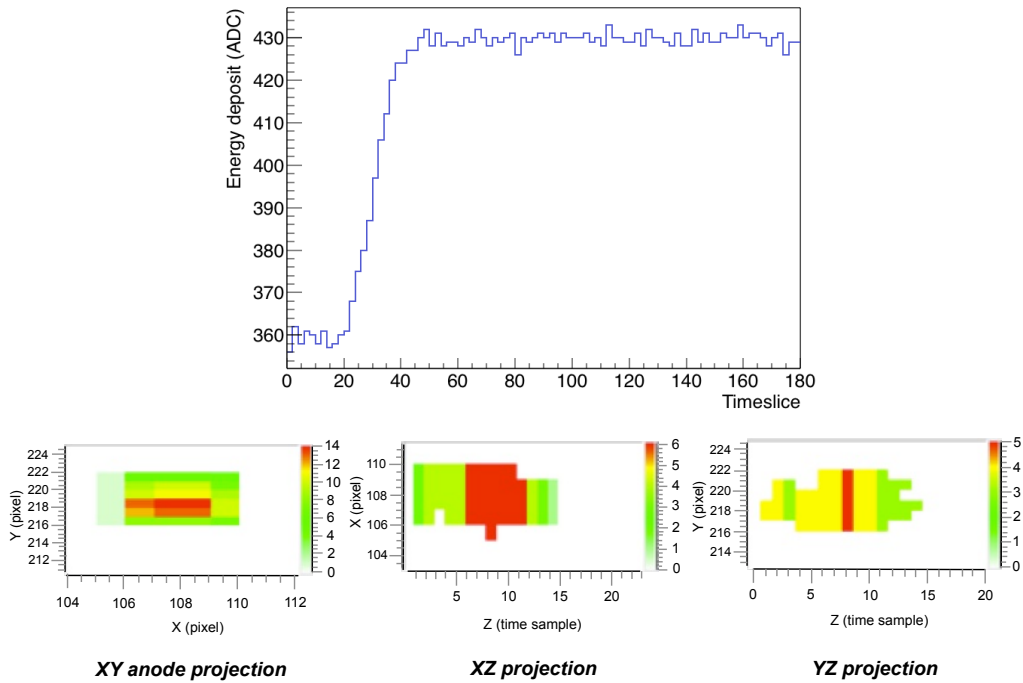

Figure 17 : A 50 keV $^4$He recoil ionization energy release profile measurement, with its 3D track, resulting from an elastic scattering with a neutron produced by the reaction T(d(220 keV),n) at the GENESIS facility.

*See online version for colors*

A selection of the events leaving clear and dense tracks in the gaseous active volume represents an excellent discrimination of the events coming from the nuclear recoils produced by elastic scattering with fast neutrons with respect to those resulting from electrons produced by gamma rays interactions [ref.16].

### 5.4 Selection of tracks contained in the gaseous active volume

As shown previously by Geant4 simulations (cf. § 5.3.2.1), some nuclear recoil tracks go outside of the gaseous active volume, or enter to this volume when mainly produced on the chamber or field cage walls. All these tracks projected on the anode have the specificity to hit the edges of the pixelated anode, which creates an over-density of events on the volume periphery (see Figure 18). These events can be rejected in setting a condition on the coordinates of the first step of the track, in order to discriminate the anode periphery.

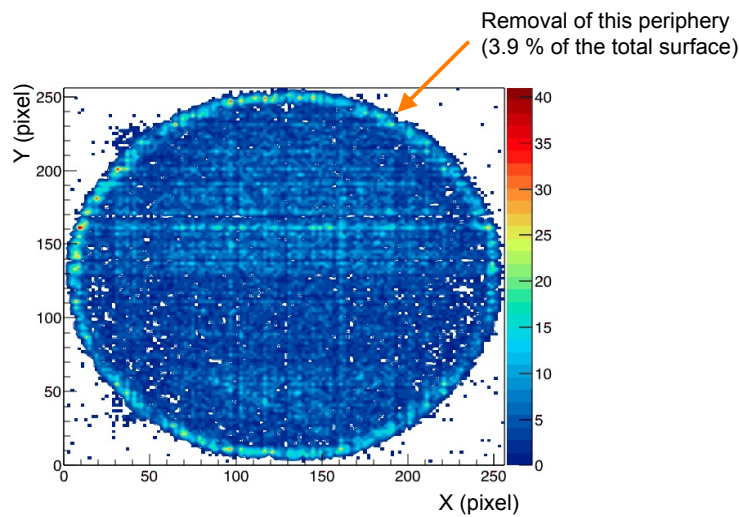

Figure 18 : First step projection on the anode of nuclear recoils' tracks, resulting from elastic scattering with neutrons produced by the reaction T(d(220 keV),n) on GENESIS.

*See online version for colors*

### 5.5 Selection of $^4$He recoils

The $^4$He branch (E) has to be selected to reject protons, $^{12}$C and $^{16}$O. Besides, we have to reject $^4$He that go outside of the active volume along the chamber longitudinal axis, since they are not reliable for the neutron spectrum calculation, because of their incomplete track and energy released in the active volume.

All these discriminations are based on the analysis of the energy released in ionization normalized to the recoil track length as a function of ionization energy, as shown in Figure 19.

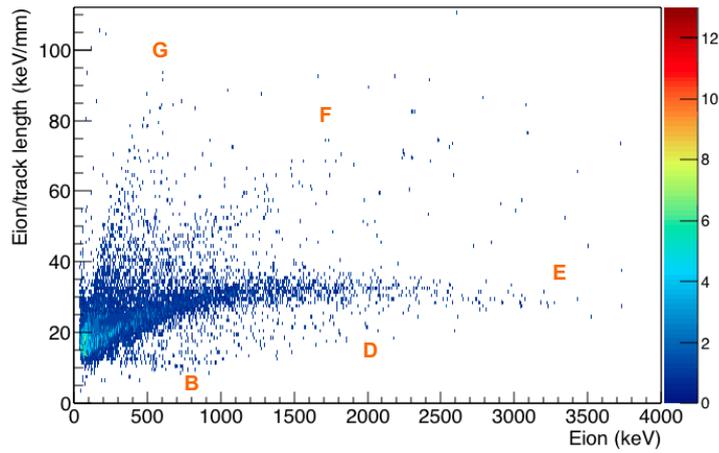

Figure 19 : Mimac-FastN data in the perpendicular configuration showing the ionization energy normalized by the track length as a function of ionization energy, for nuclear recoils resulting from elastic scattering with neutrons produced by the reaction T(d(220 keV),n) at GENESIS facility. This plot was drawn after discrimination of tracks going outside of the active volume through its lateral limits and rejection of tracks with holes (with more than 1% of empty timeslices).

*See online version for colors*

Compared to Figure 12, branch A does not appear on this plot, since its related proton tracks go outside of the volume and have characteristics of pixel density leading to their discrimination. Structure D consists of $^4$He recoils that go outside of the volume along the chamber longitudinal axis. Branch F is the branch merging $^{12}$C and $^{16}$O recoils. A new branch emerges from the enhancement due to the normalization with the track length, branch G, that consists of CO recoils, that have track lengths of roughly 5 mm for released ionization energy of 250 keV. The Figure 20 shows such a track of CO recoil. These CO molecules may have for origin a dissociation of the $CO_2$ molecules by the ionizing radiation inside the chamber during the measurements.

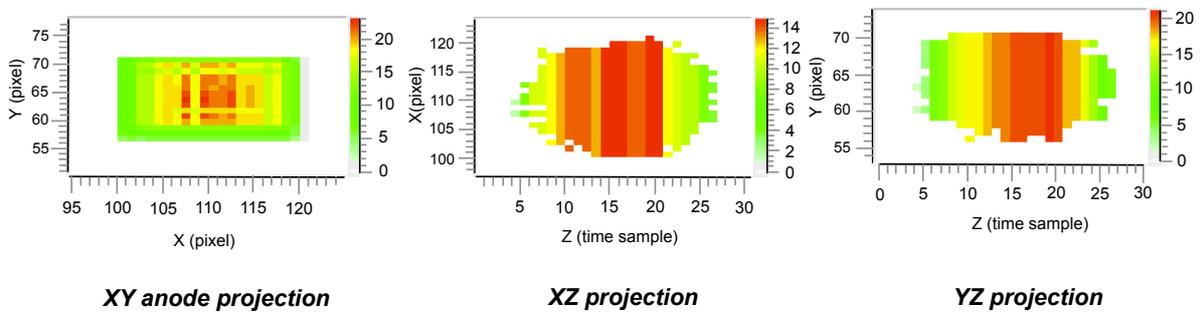

**XY anode projection**　　　　**XZ projection**　　　　**YZ projection**

Figure 20 : A 250 keV CO recoil track resulting from an elastic scattering with a neutron produced by the reaction T(d(220 keV),n) at GENESIS.

*See online version for colors*

Some of these molecular recoils are splited along their path.

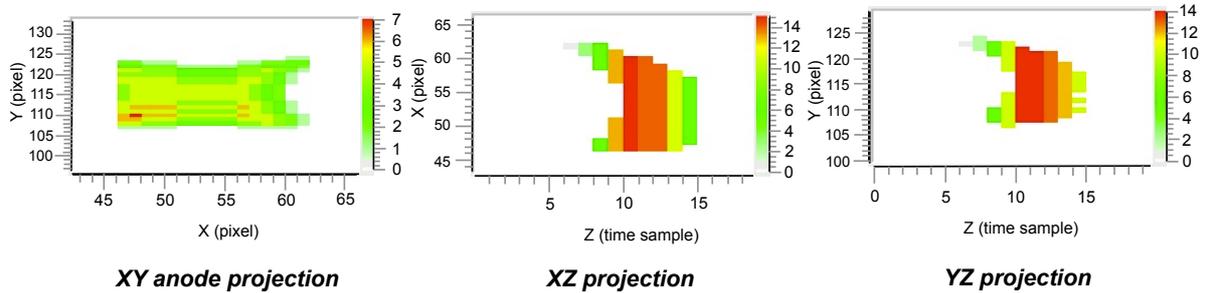

**XY anode projection**     **XZ projection**     **YZ projection**

Figure 21 : A splited CO molecular recoil track

*See online version for colors*

Second order polynomial fits of the main $^4$He recoil branch (branch E) are done down to the Flash-ADC energy threshold, to keep only recoils useful to neutron spectrum reconstruction. The proportion of remaining carbon and oxygen nuclear recoils will be estimated in our future paper dedicated to the uncertainties and energy resolution study.

## 5.6   $^4$He recoil track direction determination

Having a time sampling of 25 ns, Mimac-FastN gives access to a high-resolution profile of the charges deposited in ionization for each event. Figure 22 shows an example of such a charge profile lasting 300 timeslices, each bin width (timeslice) representing 25 ns.

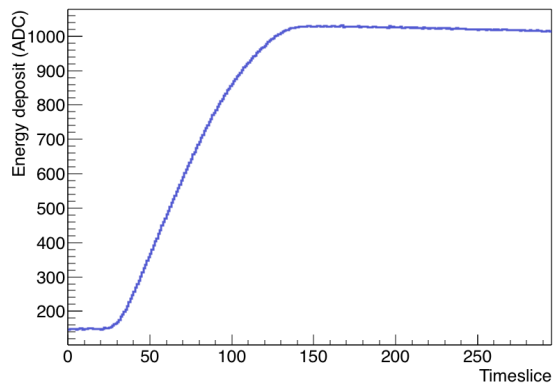

Figure 22 : Charge profile integration on the grid of an $^4$He recoil, resulting from an elastic scattering with a neutron of 15 MeV. This profile is composed of 300 time samples of 25 ns collected in the time range of 7.5 µs.

This charge profile can be analyzed event by event to detect any specificity in the amount of charges deposited as a function of time. The amount of charges expected per time sample will be higher under the Bragg peak. At energies lower than 1 MeV for $^4$He, the Bragg peak is roughly located in the first half of the track, producing more charges in one half of the charge profile compared to the other one. In such a way, a symmetry analysis of the charge profile with respect to its middle point, can be performed as illustrated in Figure 23, to find the Bragg peak location for which the maximum number of primary electrons have been produced.

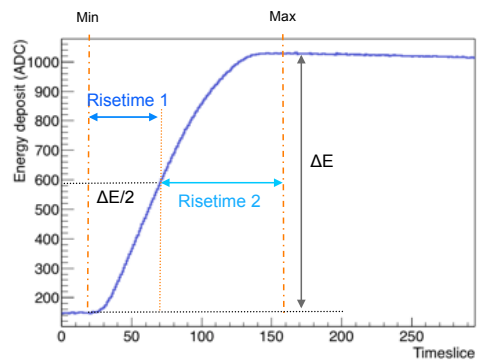

Figure 23 : Definition of risetimes 1 and 2 on the charge profile of an $^4$He recoil.

*See online version for colors*

The comparison of risetime$_1$ versus risetime$_2$ measurements constitutes an observable to define the track direction. The Figure 24 shows a plot of risetime$_2$ as a function of risetime$_1$ for $^4$He recoils resulting from elastic scattering with neutrons of 3 MeV, in a perpendicular configuration of the beam. This plot reveals two distinct branches that are assigned to each track direction.

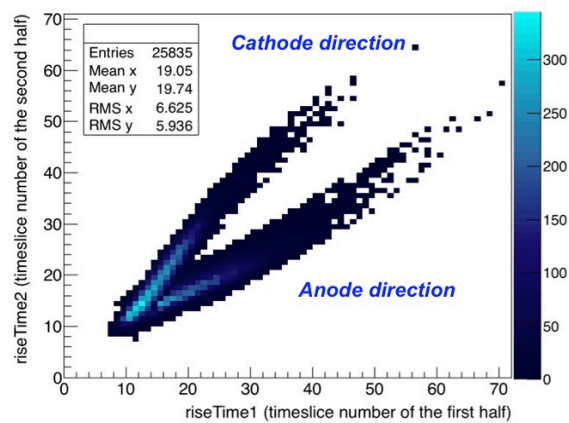

Figure 24 : Comparison of risetime$_1$ and risetime$_2$ measurements for $^4$He recoils resulting from elastic scattering with neutrons of 3 MeV, in a perpendicular configuration with respect to the beam.

*See online version for colors*

The Figure 25 shows the first and last pixels of the tracks projected on the pixelated anode, for the selection of the lower branch of the plot of risetime$_2$ as a function of risetime$_1$, assigned to the tracks directed towards the anode. The Figure 26 shows the same plots for the selection of the upper branch, assigned to the tracks directed towards the cathode. For all these plots, scattering angles below 40° have been selected highlighting the forward scattering.

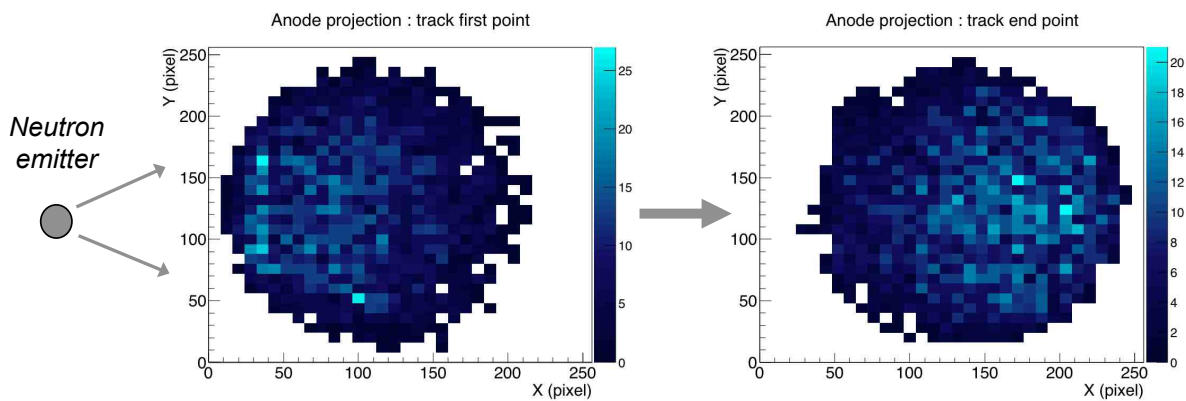

Figure 25 : Position of the first and last point of the $^4$He recoil tracks, projected on the pixelated anode, for the tracks directed towards the anode.

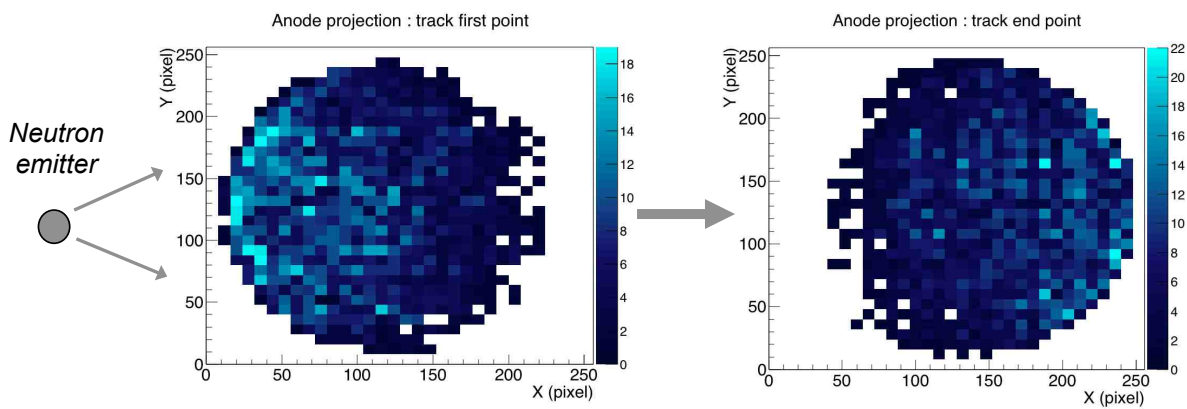

Figure 26 : Position of the first and last point of the $^4$He recoil tracks, projected on the pixelated anode, for the tracks directed towards the cathode.

Figure 25 and Figure 26 show the coordinates of the extremity points of the $^4$He recoil tracks being consistent with the position of the neutron emitter and the forward scattering of the nuclear recoil, whether the recoil is directed towards the cathode or the anode. For the selection of the events directed towards the cathode, the interaction point (first point of the track) is the first time-slice read on the anode, whereas for events directed towards the anode, the first time-slice read on the anode corresponds to the last point of the nuclear recoil track.

The measured ionization energy loss per time-slice can be related to the measured nuclear recoil path length per time-slice, through the synchronization of the readouts on the grid and on the pixelated anode. That means that the Bragg peak location can be found for each event with a precision of a time-slice.

The temporal position of the Bragg peak is plotted in Figure 27 for tracks directed towards the anode, and towards the cathode. This plot represents, for each event, the time-slice for which the Bragg peak is located, normalized to the charge collection duration, as read on the anode.

For the tracks directed towards the cathode, the Bragg peak is located at 40 % of the time development of the track on average. For the tracks directed towards the anode, it is located at 60 % on average in the reading direction on the anode, so at 40 % from the track interaction point.

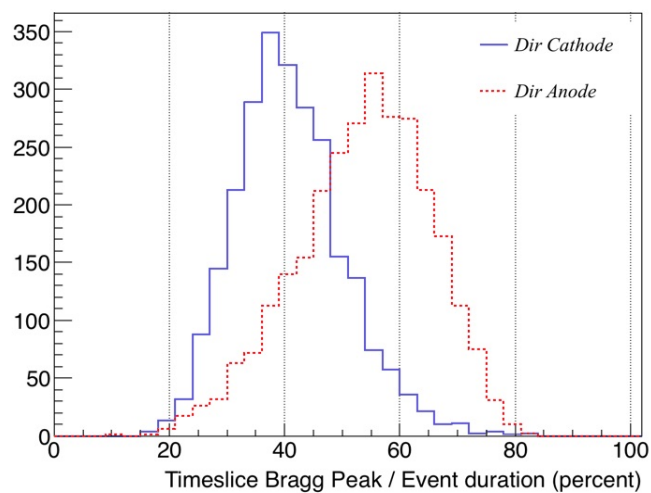

Figure 27 : Position of the Bragg peak for $^4$He recoils resulting from elastic scattering with neutrons of 3 MeV issued from D(d(220 keV),n) on GENESIS, for tracks directed towards the anode or the cathode.

*See online version for colors*

A dedicated paper on these head-tail MIMAC measurements on nuclear recoil tracks is in preparation.

## 6    Experimental results from D(d(220)keV,n) reaction

Spectrometry of the neutrons produced by the reaction D(d(220 keV,n) has been performed at the GENESIS facility [ref.13], with Mimac-FastN placed at 0° degree with respect to the deuteron beam axis. In this configuration, the D(d(220 keV,n) nuclear reaction produces neutrons of 3.1 MeV.
The target is a solid target, composed of titanium loaded with deuterium, and evaporated on a 3 mm thick copper backing [ref.17]. The measurement has been done in one hour, with a deuteron current of 100 µA, in the configuration with the longitudinal axis of the detector perpendicular to the beam axis. The detector was positioned at 1 meter from the target. In this directional measurement, we consider the target location as the neutron source position. The spectrometer Mimac-FastN was filled with a gas mixture of 95 % of $^4$He and 5 % of $CO_2$ at 700 mbar.
All the discrimination principles presented previously in the paragraph entitled "Data analysis strategy" have been applied to the data. Besides, fast neutrons can be scattered on the concrete walls of the bunker. The degrees of freedom brought by the 3D geometry have been explored to discriminate part of these scattered neutrons on the walls of the facility. The resulting neutron spectrum is normalized to the acquisition integrated time corrected from the dead time, and to the detection efficiency calculated by simulation with Geant4, which embeds the elastic scattering cross section (see Figure 28).

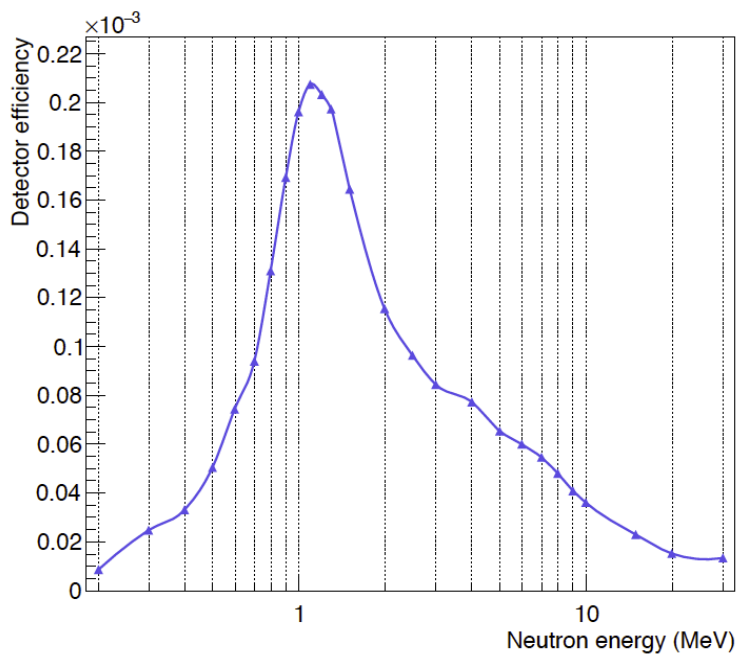

Figure 28 : detection efficiency curve calculated by simulation with Geant4

The final neutron energy spectrum is plotted in Figure 29.

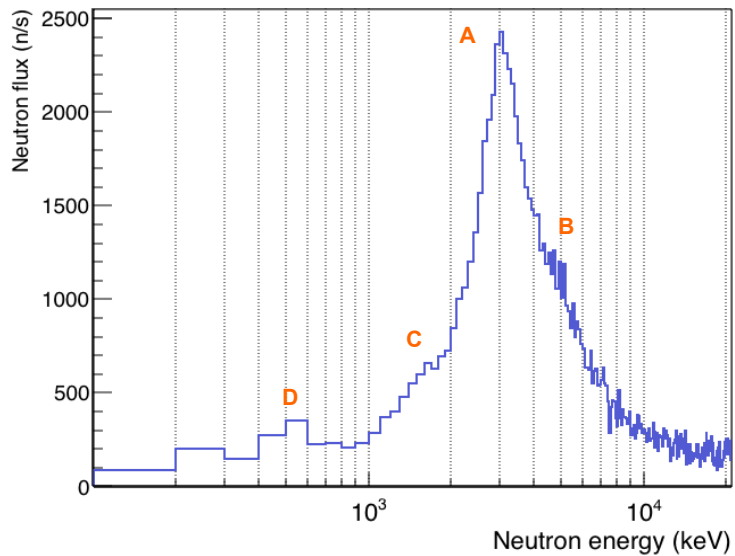

Figure 29 : Measured neutron spectrum with the reaction D(d(220 keV),n) at GENESIS, with a binning of 120 keV/bin.

The measured neutron spectrum reveals a polyenergetic spectrum with 4 structures. This is explained by the target composition and the interaction of deuterons with all the components of the target.

The peak **A** is centered around 3.1 MeV, that is the mean energy expected of neutrons at 0° resulting from the reaction D(d(220 keV,n). The shoulder **B** is produced by neutrons resulting from $^{63}$Cu(d,n)$^{64}$Zn and $^{65}$Cu(d,n)$^{66}$Zn with energies between 5.5 MeV and 6.8 MeV, and from $^{48}$Ti(d,n)$^{49}$V with energies around 4.5 MeV. The shoulder **C** results from $^{64}$Zn(d,n)$^{65}$Ga with energies around 1.8 MeV. The structure **D** is a contribution of the residual scattered neutrons on the walls of the accelerator bunker.

These measurements reveal that cross sections of low energy deuterons on copper and zinc are not negligible, despite the lack of data reported in the literature on this subject. This measured neutron spectrum is associated to the measured angular distribution of $^4$He recoils related to the incident neutron direction. This measured angular distribution is presented in Figure 30 and shows the same feature as the simulated distribution given by Geant4 in Figure 3, namely, a higher probability of elastic scattering between 50° and 75°, which confirms the Geant4 simulation features.

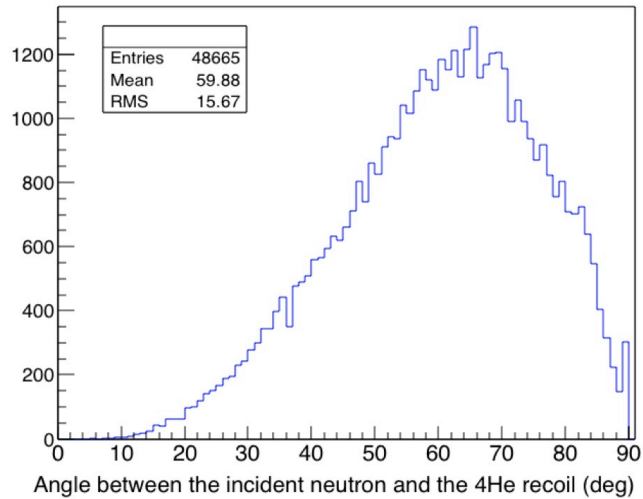

Figure 30 : Angular distribution of $^4$He recoils resulting from elastic scattering with neutrons, from the reaction D(d(220 keV),n) at GENESIS.

## 7 Experimental results from T(d(220)keV,n) reaction

At the same facility and with the same experimental set-up, spectrometry of the neutrons produced by the reaction T(d(220 keV,n) has been performed. At $0°$ degrees with respect to the deuteron beam axis, the T(d(220 keV,n) nuclear reaction produces neutrons of 15.1 MeV.

The target has the same structure as the one described previously with the deuterium loading replaced by a tritium loading.

The measurement has been done in 1 hour, with a deuteron current of 5 µA. The neutron spectrometer longitudinal axis is perpendicular to the beam axis. It was located at 1.7 meters from the target, and the gas mixture was the same as the one used for the reaction D(d(220 keV),n) experiment. Figure 2 shows a picture of the experimental set-up.

The Figure 31 shows the reconstructed neutron spectrum.

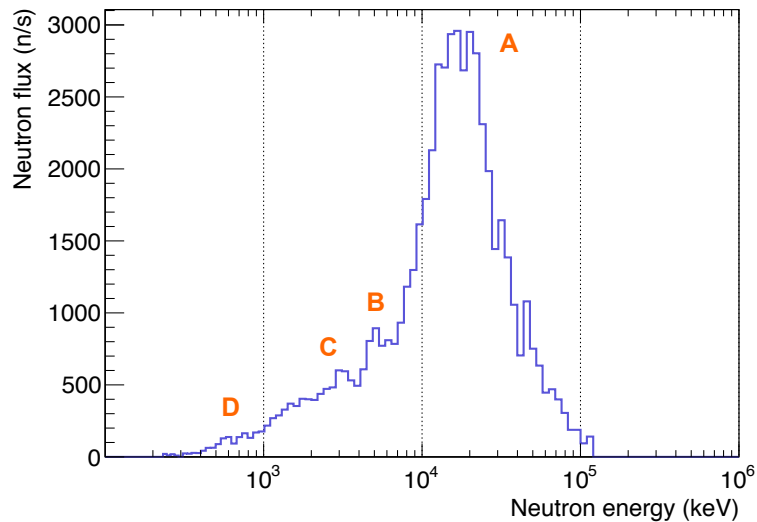

Figure 31 : Measured neutron spectrum with the reaction T(d(220 keV),n) at GENESIS, with a binning of 320 keV/bin.

As in the precedent experiment, a polyenergetic spectrum is observed.

The peak **A** is the expected production of neutrons of 15.1 MeV resulting from the reaction T(d(220 keV,n). The peak **B** is populated by neutrons resulting from $^{63}$Cu(d,n)$^{64}$Zn, $^{65}$Cu(d,n)$^{66}$Zn and $^{48}$Ti(d,n)$^{49}$V. The peak **C** results from D(d,n) reactions following the implantation of part of the incident deuterons into the target backing. The peak **D** is a contribution of the residual scattered neutrons on the walls of the accelerator bunker.

This neutron spectrum is associated to the angular distribution of $^4$He recoils related to the incident neutron direction. This angular distribution is presented in Figure 32, and shows the same feature as the simulated distribution given by Geant4 in Figure 3, namely, a higher probability of emission of the $^4$He recoil around 70°.

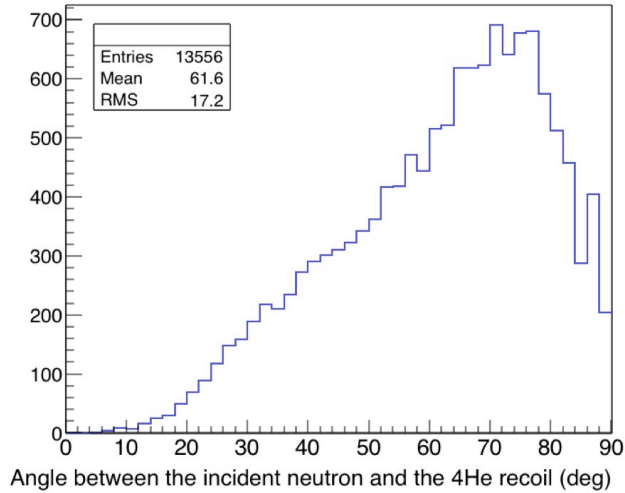

Figure 32 : Angular distribution of $^4$He recoils resulting from elastic scattering with neutrons, from the reaction T(d(220 keV),n) at GENESIS.

As estimated by Geant4 simulations shown in Figure 3, the measurements confirm that the higher the neutron energy is, the higher the probability that the $^4$He recoil will be scattered at an angle greater than 45$^o$ with respect to the incident neutron direction.

## 8    Discussion on additional work to achieve neutron spectroscopy

In the previous paragraphs, we have presented neutron spectra of mono-energetic neutron fields, to demonstrate the discrimination capabilities and the method performance for describing structures in the neutron spectra in a large energy range, with the gas mixture proposed. In a future work, we will focus on a more quantitative study on uncertainties estimation of the neutron energy reconstruction, and on the neutron fluence by comparison with a reference detector, as it was started in the frame of two PhD thesis in collaboration with the LMDN (IRSN) laboratory thesis [ref.6 and 7].

## 9    Conclusion

In the present paper, we have described the ability of Mimac-FastN to measure mono-energetic neutron spectra at 3 MeV and 15 MeV. Using the same gas mixture, with a threshold on the neutron energy as low as 200 keV, this directional fast neutron spectrometer gives a complete polyenergetic neutron spectrum exploring the material and eventual pollutions of the target or neutron sources.

This ability to provide polyenergetic neutron spectrum has already been applied to characterize the angular distribution of fast neutrons produced in a nuclear reaction proposed for a radiotherapy called Accelerator-Based Boron Neutron Capture Therapy (AB-BNCT) [ref.18].

With its high spatial resolution 3D track reconstruction associated with its large adjustable measuring range, Mimac-FastN is a versatile instrument that opens new fields for directional neutron spectrometry, for applications such as nuclear matter characterization, detection of target pollution, nuclear cross section measurements or monitoring the neutron production for radioprotection purposes. As the measurement is directional, the neutron source position has to be known, or previously found with a rough estimate of its nature using a reverse algorithm.

Besides this study, preliminary measurements performed at CERF (CERN) [ref.19] with this instrument have recently shown a good potential for spectrometry at neutron energies as high as 200 MeV, a range covering atmospheric neutron production assuming that the neutron directions were parallel to the anode plane, as well as for high energy neutron monitoring.

## 10  Acknowledgments

This work has been funded by the "Prematuration" program of CNRS, by the LabEx Enigmass, and by Linksium SATT (Technology Transfer Accelerator Office).

We thank the LPSC accelerator team for the operation of the GENESIS facility labeled CNRS for their support during all the performed experiments.